\documentclass[twocolumn]{aastex62}

\usepackage{bm}
\usepackage{amsmath}

\definecolor{MyDarkBlue}{rgb}{0,0.08,0.5}
\definecolor{MyDarkRed}{rgb}{0.7,0.02,0.02}

\received{October 17, 2018}
\revised{February 19, 2019}
\accepted{February 21, 2019}
\submitjournal{ApJ}

\shorttitle{Tensile Strength of Porous Dust Aggregates} 
\shortauthors{Tatsuuma et al.}

\begin{document}

\title{Tensile Strength of Porous Dust Aggregates}

\author[0000-0003-1844-5107]{Misako Tatsuuma}
\email{misako.tatsuuma@nao.ac.jp}
\affil{Department of Astronomy, Graduate School of Science, The University of Tokyo, 7-3-1 Hongo, Bunkyo-ku, Tokyo 113-0033, Japan}
\affil{Division of Theoretical Astronomy, National Astronomical Observatory of Japan, 2-21-1 Osawa, Mitaka, Tokyo 181-8588, Japan}

\author[0000-0003-4562-4119]{Akimasa Kataoka}
\affil{Division of Theoretical Astronomy, National Astronomical Observatory of Japan, 2-21-1 Osawa, Mitaka, Tokyo 181-8588, Japan}

\author{Hidekazu Tanaka}
\affil{Astronomical Institute, Graduate School of Science, Tohoku University, 6-3 Aramaki, Aoba-ku, Sendai 980-8578, Japan}

\begin{abstract} 

Comets are thought to have information about the formation process of our solar system.
Recently, detailed information about comet 67P/Churyumov-Gerasimenko has been found by a spacecraft mission Rosetta.
It is remarkable that its tensile strength was estimated.
In this paper, we measure and formulate the tensile strength of porous dust aggregates using numerical simulations, motivated by porous dust aggregation model of planetesimal formation.
We perform three-dimensional numerical simulations using a monomer interaction model with periodic boundary condition.
We stretch out a dust aggregate with a various initial volume filling factor between $10^{-2}$ and 0.5.
We find that the tensile stress takes the maximum value at the time when the volume filling factor decreases to about a half of the initial value.
The maximum stress is defined to be the tensile strength.
We take an average of the results with 10 different initial shapes to smooth out the effects of initial shapes of aggregates.
Finally, we numerically obtain the relation between the tensile strength and the initial volume filling factor of dust aggregates.
We also use a simple semi-analytical model and successfully reproduce the numerical results, which enables us to apply to a wide parameter range and different materials.
The obtained relation is consistent with previous experiments and numerical simulations about silicate dust aggregates.
We estimate that the monomer radius of comet 67P has to be about 3.3--220 $\mathrm{\mu m}$ to reproduce its tensile strength using our model.

\end{abstract}

\keywords{planets and satellites: formation --- protoplanetary disks --- methods: numerical --- methods: analytical}

\section{Introduction} \label{sec:intro}

Planetesimal formation is one of the most important and unsolved issues of planet formation theory.
In protoplanetary disks, sub-$\mathrm{\mu}$m-sized dust grains are believed to coagulate, settle to the disk midplane as they grow, and form km-sized planetesimals.
There are several scenarios about the planetesimal formation such as gravitational instability \citep[e.g.,][]{Goldreich1973}, streaming instability \citep[e.g.,][]{Youdin2005,Johansen2007,Johansen2011}, and direct coagulation.
In the direct coagulation scenario, dust grains grow larger by pairwise collisions.
Recently, it has been proposed that dust grains become not compact but porous by pairwise collisions, and properties of these fluffy dust aggregates have been investigated theoretically and experimentally \citep[e.g.,][]{Kozasa1992,Ossenkopf1993,Dominik1997,Blum2000,Wada2007,Wada2008,Suyama2008}.
The sub-$\mathrm{\mu}$m-sized constituent grains are called monomers.
Finally, it is found that planetesimals form via direct coagulation \citep[e.g.,][]{Okuzumi2012,Kataoka2013L}.

In recent years, physical properties of comets have been investigated by observation and exploration.
Comets are the most primitive bodies in our solar system and are thought to be leftover planetesimals.
In 2014, a spacecraft Rosetta reached comet 67P/Churyumov-Gerasimenko (hereinafter 67P).
This mission was the first one to orbit and land onto a comet.
There are many unexpected results about 67P \citep[e.g.,][]{Fulle2016}, and especially it is remarkable that its tensile strength was estimated.
The tensile strength of 67P for its surface is 3--150 Pa \citep{Groussin2015,Basilevsky2016}, while for bulk comet 10--200 Pa \citep{Hirabayashi2016}.
This tensile strength depends on composition and formation process of comets, i.e., planetesimals.

There are several experimental studies about the tensile strength of dust aggregates.
\citet{Blum2004} directly measured the tensile strength of dust aggregates whose volume filling factors are 0.2 and 0.54.
They used dust aggregates consisted of monodisperse silica (SiO$_2$) spheres with $0.76\mathrm{\ \mu m}$ radius.
In their experiments, a mm-sized dust aggregate was attached to two plates at its top and bottom, and then the two plates were pulled apart.
\citet{Blum2006} conducted the same experiments using dust aggregates which have volume filling factors of 0.23, 0.41, and 0.66.
In addition to the monodisperse spherical silica monomers, they used irregularly shaped diamond monomers with a narrow size distribution and irregular silica monomers with a wide size distribution.
\citet{Meisner2012} used dust aggregates consisted of quartz (crystallized SiO$_2$) monomers with a size range from 0.1 $\mathrm{\mu m}$ to 10 $\mathrm{\mu m}$ and measured the tensile strength using the Brazilian disc test \citep[e.g.,][]{Li2013}.
\citet{Gundlach2018} also performed the Brazilian disc test to measure the tensile strength of dust aggregates composed of polydisperse spherical ice (H$_2$O) monomers and monodisperse spherical silica monomers.
They used silica monomers whose radii are $0.15\mathrm{\ \mu m}$, $0.50\mathrm{\ \mu m}$, and $0.75\mathrm{\ \mu m}$ to investigate the monomer radius dependence.
Moreover, they succeeded in measuring the tensile strength of ice dust aggregates whose monomer radius is 2.4 $\mathrm{\mu m}$ on average.

On the other hand, there is only one numerical study about the tensile strength of dust aggregates.
\citet{Seizinger2013} performed three-dimensional simulations to reproduce the experimental results by \citet{Blum2004} and \citet{Blum2006}.
They used dust aggregates whose volume filling factor ranges from 0.15 to 0.6 and monomers are silicate spheres with 0.6 $\mathrm{\mu m}$ radius.
In their simulations, a $\mathrm{\mu m}$-sized cubic aggregate was attached to two plates, which is the same as previous experiments except for the size of dust aggregates; a mm-sized dust aggregate was used in the previous experiments.
The interaction between two monomers is mainly based on \citet{Dominik1997}.
In addition, they introduced the rolling and sliding modifiers to make numerical simulations correspond with experimental results \citep{Seizinger2012}.
To avoid for monomers being peeled off the plate, they also used artificial adhesion force as ``gluing effect."
Although their results correspond well with the laboratory ones, the influences of their artificial adhesion force and small aggregates should also be checked.
They also obtained a fitting formula of the tensile strength as a function of the filling factor of dust aggregates.
However, their formula does not include the dependence on the monomer size and material.

In this work, we numerically investigate the tensile strength of dust aggregates composed of single-sized spherical monomers.
In the previous works, a dust aggregate was attached to two plates, and then they were pulled apart \citep{Blum2004,Blum2006,Seizinger2013}.
The size of the used dust aggregates ranges from $\mathrm{\mu m}$ to mm, while planetesimals are km-sized.
To unravel the planetesimal formation mechanism, it is important to investigate the tensile strength of dust aggregates whose size is larger than km.
Therefore, we use the periodic boundary condition to remove effects of plates.
Moreover, we perform simulations using dust aggregates whose volume filling factors are lower than those of the previous works.
Then, we find a power-law relation between the tensile strength and initial volume filling factor of dust aggregates whose filling factors range from $10^{-2}$ to 0.5.
We will also construct a theoretical model to explain the power-law dependence.
This model reproduces the dependence on all material parameters in our simulations.

This paper is constructed as follows.
In Section \ref{sec:setting}, we describe settings of our simulations, which include the model of monomer interactions, initial dust aggregates, periodic boundary condition, how to measure the tensile strength without plates, and an overview of our simulations.
Then, we summarize our results of fiducial runs and the investigation of parameter dependence in Section \ref{sec:results}.
There are three numerical parameters: the number of monomers, boundary velocity, and the strength of the damping force, while four physical parameters: the initial volume filling factor, monomer radius, surface energy, and the critical rolling displacement.
We also find an analytical expression of the tensile strength of dust aggregates, compare our results with previous experiments \citep{Blum2004,Blum2006,Gundlach2018} and numerical simulations \citep{Seizinger2013}, and apply the analytical expression to comet 67P in Section \ref{sec:discuss}.
Finally, we conclude this work and discuss future works in Section \ref{sec:conclusions}.

\section{Simulation Settings} \label{sec:setting}

We perform three-dimensional numerical simulations to measure the tensile strength of dust aggregates consisting of spherical monomers.
In this section, we describe settings of our simulations.
First, we introduce the monomer interaction model based on \citet{Dominik1997} and \citet{Wada2007} in Section \ref{subsec:interact}.
In Section \ref{subsec:damping}, we explain the damping force in the normal direction.
The initial conditions of our simulations are statically and isotropically compressed dust aggregates investigated by \citet{Kataoka2013}, which is described in Section \ref{subsec:initial}.
At the boundaries of the calculation box, we set moving periodic boundaries in the $x$-axis direction and fixed periodic boundaries in the $y$- and $z$-axis directions, which is explained in Section \ref{subsec:boundary}.
Thus, we can simulate one-direction stretching of dust aggregates.
The details of the calculation method of tensile stress, which is the same as molecular dynamics, is summarized in Section \ref{subsec:stress}.
In Section \ref{subsec:overview}, we describe an overview of our simulations.

\subsection{Monomer Interaction Model} \label{subsec:interact}

We calculate interactions of each connection between two monomers using the theoretical model by \citet{Dominik1997} and \citet{Wada2007}.
Based on the JKR theory \citep{Johnson1971} and the following studies by \citet{Dominik1995,Dominik1996}, \citet{Dominik1997} carried out two-dimensional simulations of monomer interactions.
To expand into three-dimensional simulations, \citet{Wada2007} tested their recipe, and then \citet{Wada2008} conducted three-dimensional simulations of dust aggregate collisions.
In the model, there are four kinds of interactions named normal (sticking and breaking), sliding, rolling, and twisting.
The material parameters needed to describe the interactions are the monomer radius $r_0$, material density $\rho_0$, surface energy $\gamma$, Poisson's ratio $\nu$, Young's modulus $E$, and the critical rolling displacement $\xi_\mathrm{crit}$.
These parameters of ice and silicate are listed in Table \ref{tab:param}.
To compare our results with those by \citet{Seizinger2013}, we set the same values for silicate.

If a rolling displacement exceeds the critical one $\xi_\mathrm{crit}$, a monomer begins to roll inelastically.
The critical rolling displacement has different values between the theoretical one \citep[$\xi_\mathrm{crit}=2$ \AA,][]{Dominik1997} and the experimental one \citep[$\xi_\mathrm{crit}=32$ \AA,][]{Heim1999}.
We adopt $\xi_\mathrm{crit}=8$ \AA\ as a fiducial value and investigate the dependence of our results on $\xi_\mathrm{crit}$ in Section \ref{subsec:phyparamdepend}.

The rolling energy $E_\mathrm{roll}$ needed to rotate a monomer around its connection point by 90$^\circ$ is described as
\begin{eqnarray}
E_\mathrm{roll} &=& 12\pi^2\gamma R\xi_\mathrm{crit}\nonumber\\
&=& 6\pi^2\gamma r_0\xi_\mathrm{crit}\nonumber\\
&\sim& 4.7\times10^{-16}\left(\frac{\gamma}{100\mathrm{\ mJ\ m^{-2}}}\right)\nonumber\\
&&\times\left(\frac{r_0}{0.1\mathrm{\ \mu m}}\right)\left(\frac{\xi_\mathrm{crit}}{8\textrm{\ \AA}}\right)\mathrm{\ J},
\end{eqnarray}
where $R$ is the reduced monomer radius \citep{Wada2007}.
The reduced radius $R$ of monomer radii $r_1$ and $r_2$ is defined as
\begin{equation}
\frac{1}{R} = \frac{1}{r_1}+\frac{1}{r_2}.
\end{equation}
Here, the reduced monomer radius is $R=r_0/2$ because we assume no size distribution of monomers.

In our simulations, the maximum force needed to separate two sticking monomers (breaking) is
\begin{eqnarray}
F_\mathrm{c}&=&3\pi\gamma R\nonumber\\
&\sim&4.7\times10^{-8}\left(\frac{\gamma}{100\mathrm{\ mJ\ m^{-2}}}\right)\left(\frac{r_0}{0.1\mathrm{\ \mu m}}\right)\mathrm{\ N}.
\label{eq:Fc}
\end{eqnarray}

\begin{table}
\centering
\caption{Material parameters of ice \citep{Israelachvili1992,Dominik1997}.
The parameters of silicate are selected according to \citet{Seizinger2013}.
}
\label{tab:param}
\begin{tabular}{ccc}
\hline\hline
Material & Ice & Silicate\\\hline
Monomer radius $r_0$ [$\mathrm{\mu m}$] & 0.1 & 0.6\\
Material density $\rho_0$ [g cm$^{-3}$] & 1.0 & 2.65\\
Surface energy $\gamma$ [mJ m$^{-2}$] & 100 & 20\\
Poisson's ratio $\nu$ & 0.25 & 0.17\\
Young's modulus $E$ [GPa] & 7 & 54 \\
Critical rolling displacement $\xi_\mathrm{crit}$ [\AA] & 8 & 20 \\
\hline
\end{tabular}
\end{table}

\subsection{Damping Force in Normal Direction} \label{subsec:damping}

The force in the normal direction induces oscillation at each connection between two monomers.
In reality, the oscillation would attenuate because of viscoelasticity or hysteresis of monomers \citep{Greenwood2006,Tanaka2012}.
Therefore, we add an artificial damping force in the normal direction \citep{Suyama2008,Paszun2008,Seizinger2012,Kataoka2013}.
The dependence of our results on the damping force is investigated in Section \ref{subsec:numparamdepend}.

We describe the damping force as follows.
In the case that two contacting monomers have their position vectors $\bm{x}_1$ and $\bm{x}_2$, and velocities $\bm{v}_1$ and $\bm{v}_2$, respectively, the contact unit vector $\bm{n}_\mathrm{c}$ is defined as
\begin{equation}
\bm{n}_\mathrm{c} = \frac{\bm{x}_1-\bm{x}_2}{|\bm{x}_1-\bm{x}_2|}
\end{equation}
\citep{Wada2007}.
The damping force applied to each monomer is introduced as
\begin{equation}
\bm{F}_\mathrm{damp} = -k_\mathrm{n}\frac{m_0}{t_\mathrm{c}}(\bm{n}_\mathrm{c}\cdot\bm{v}_\mathrm{r})\bm{n}_\mathrm{c},
\end{equation}
where $k_\mathrm{n}$ is the damping coefficient, $m_0$ is the monomer mass, $t_\mathrm{c}$ is the characteristic time, and $\bm{v}_\mathrm{r}$ is the relative velocity \citep{Kataoka2013}.
When we calculate the damping force experienced by the monomer ($\bm{x}_1,\bm{v}_1$), $\bm{v}_\mathrm{r}=\bm{v}_2-\bm{v}_1$ is the relative velocity of the other monomer ($\bm{x}_2,\bm{v}_2$).
We adopt $k_\mathrm{n}=0.01$ as a fiducial value.

The characteristic time is given by \citet{Wada2007} as
\begin{equation}
t_\mathrm{c} = 0.95\left(\frac{r_0^{7/6}\rho_0^{1/2}}{\gamma^{1/6}{E^\ast}^{1/3}}\right),
\end{equation}
where $E^\ast$ is the reduced Young's modulus of monomers 1 and 2 defined as
\begin{equation}
\frac{1}{E^\ast} = \frac{1-\nu_1^2}{E_1}+\frac{1-\nu_2^2}{E_2}.
\end{equation}
In our simulation, the Young's modulus $E_1=E_2=E$ and the Poisson's ratio $\nu_1=\nu_2=\nu$ are uniform.

\subsection{Initial Dust Aggregates} \label{subsec:initial}

The initial dust aggregates are statically and isotropically compressed ballistic cluster-cluster aggregations (BCCAs) investigated by \citet{Kataoka2013}.
We set these initial conditions to simulate the planetesimal formation mechanism.
The calculation boundary is treated periodically, thus we do not have to consider the aggregate radius.

\subsection{One-Direction Stretching by Moving Boundaries} \label{subsec:boundary}

We set moving boundaries in the $x$-axis direction and fixed boundaries in the $y$- and $z$-axis directions to measure the tensile strength of dust aggregates.
The initial calculation box is a cube whose length on each side is $L_0$.
The length in the $y$- and $z$-axis directions does not change, while the length in the $x$-axis direction $L_x$ increases.
Therefore, the coordinates in the $x$-, $y$-, and $z$-axis directions are $-L_x/2<x<L_x/2$, $-L_0/2<y<L_0/2$, and $-L_0/2<z<L_0/2$, respectively.

The velocity at the boundary in the $x$-axis direction $v_\mathrm{b}>0$ has to be constant and less than the effective sound speed of dust aggregates for statical stretching.
We investigate the dependence on the velocity in Section \ref{subsec:numparamdepend}.
The effective sound speed of dust aggregates $c_\mathrm{s,eff}$ is described as
\begin{equation}
c_\mathrm{s,eff} \sim \sqrt{\frac{P}{\rho}},
\label{eq:sound}
\end{equation}
where $P$ and $\rho$ are the pressure and mean internal density of dust aggregates, respectively.
Because the initial dust aggregates are statically and isotropically compressed, their pressure is given as
\begin{equation}
P = \frac{E_\mathrm{roll}}{r_0^3}\left(\frac{\rho}{\rho_0}\right)^3
\label{eq:Pcomp}
\end{equation}
\citep{Kataoka2013}.
From Equation (\ref{eq:sound}) and (\ref{eq:Pcomp}), we can obtain the effective sound speed of dust aggregates as
\begin{eqnarray}
c_\mathrm{s,eff} &\sim& \sqrt{\frac{E_\mathrm{roll}}{\rho_0r_0^3}}\frac{\rho}{\rho_0}\nonumber\\
&\sim&2.2\times10^3\left(\frac{r_0}{0.1\mathrm{\ \mu m}}\right)^{-1}\left(\frac{\rho_0}{1.0\mathrm{\ g\ cm^{-3}}}\right)^{-1/2}\nonumber\\
&&\times\left(\frac{\gamma}{100\mathrm{\ mJ\ m^{-2}}}\right)^{1/2}\left(\frac{\xi_\mathrm{crit}}{8\textrm{\ \AA}}\right)^{1/2}\phi\mathrm{\ cm\ s^{-1}},
\label{eq:sound2}
\end{eqnarray}
where $\phi=\rho/\rho_0$ is the volume filling factor of dust aggregates.
Since $v_\mathrm{b}$ is independent of time $t$, the length in the $x$-axis direction $L_x$ can be written as
\begin{equation}
L_x = L_0 + 2v_\mathrm{b}t.
\end{equation}

We treat the coordinates and velocity of a monomer across a periodic boundary as follows.
When a monomer passes the moving periodic boundary at $x=L_x/2$, its position $x$ and velocity $v_x$ are converted as
\begin{eqnarray}
x &\to& x-L_x\\
v_x &\to& v_x-2v_\mathrm{b}.
\end{eqnarray}
On the other hand, in the case of the moving periodic boundary at $x=-L_x/2$, its position and velocity are converted as
\begin{eqnarray}
x &\to& x+L_x\\
v_x &\to& v_x+2v_\mathrm{b}.
\end{eqnarray}
At $y=\pm L_0/2$ and $z=\pm L_0/2$, the coordinates of a monomer are converted similarly, but its velocity is not changed.

\subsection{Tensile Stress Measurement} \label{subsec:stress}

We calculate tensile stress in the same way as \citet{Kataoka2013} because we have no walls.
This is different from \citet{Seizinger2013}, who measured the tensile stress considering the force exerted on walls.

The tensile stress is calculated only in the $x$-axis direction as follows.
At first, we assume a virtual box, which is the same as the calculation box.
The equation of motion of the monomer $i$ in the $x$-axis direction is described as
\begin{equation}
m_0\frac{\mathrm{d}^2x_i}{\mathrm{d}t^2}=W_{x,i}+F_{x,i},
\label{eq:EOM}
\end{equation}
where $W_{x,i}$ is the force exerted from the walls of the virtual box on the monomer $i$ and $F_{x,i}$ is the total force from other monomers on the monomer $i$.
We multiply Equation (\ref{eq:EOM}) by $x_i$ and take a long-time average with the time interval $\tau$.
The left-hand side of Equation (\ref{eq:EOM}) becomes
\begin{equation}
\frac{m_0}{\tau}\int_0^\tau x_i\frac{\mathrm{d}^2x_i}{\mathrm{d}t^2}dt=\frac{m_0}{\tau}\left[x_i\frac{\mathrm{d}x_i}{\mathrm{d}t}\right]_0^\tau-\frac{m_0}{\tau}\int_0^\tau\frac{\mathrm{d}x_i}{\mathrm{d}t}\frac{\mathrm{d}x_i}{\mathrm{d}t}dt.
\label{eq:EOMleft}
\end{equation}
The first term on the right-hand side of Equation (\ref{eq:EOMleft}) becomes zero when $\tau\to\infty$.
Here, we define the long-time average as $\langle\rangle_t$ and take a summation of all monomers of Equation (\ref{eq:EOM}).
Then, Equation (\ref{eq:EOM}) can be written as
\begin{equation}
\left\langle\sum_{i=1}^N\frac{m_0}{2}\left(\frac{\mathrm{d}x_i}{\mathrm{d}t}\right)^2\right\rangle_t = -\frac{1}{2}\left\langle\sum_{i=1}^Nx_iW_{x,i}\right\rangle_t -\frac{1}{2}\left\langle\sum_{i=1}^Nx_iF_{x,i}\right\rangle_t.
\label{eq:EOM2}
\end{equation}
The left-hand side of Equation (\ref{eq:EOM2}) can be defined as
\begin{equation}
K_x = \left\langle\sum_{i=1}^N\frac{m_0}{2}\left(\frac{\mathrm{d}x_i}{\mathrm{d}t}\right)^2\right\rangle_t,
\end{equation}
which is the time-averaged kinematic energy in the $x$-axis direction of all monomers.
The first energy term on the right-hand side of Equation (\ref{eq:EOM2}) is related to the tensile stress in the $x$-axis direction $P_x$.
Since the virtual box is the same as the calculation box, we obtain
\begin{equation}
\left\langle\sum_{i=1}^Nx_iW_{x,i}\right\rangle_t = L_xP_xL_0^2 = P_xV,
\end{equation}
where $V=L_xL_0^2$ is the volume of the calculation box.
Therefore, Equation (\ref{eq:EOM2}) gives an expression of the tensile stress $P_x$ as
\begin{equation}
P_x = -\frac{2K_x}{V} - \frac{1}{V}\left\langle\sum_{i=1}^Nx_iF_{x,i}\right\rangle_t.
\label{eq:Px}
\end{equation}
The total force from other monomers on the monomer $i$ can be described as
\begin{equation}
F_{x,i} = \sum_{j\neq i}f_{x,i,j},
\end{equation}
where $f_{x,i,j}$ is the force from the monomer $j$ on the monomer $i$ in the $x$-axis direction.
Thus, Equation (\ref{eq:Px}) can be written as
\begin{equation}
P_x = -\frac{2K_x}{V} - \frac{1}{V}\left\langle\sum_{i<j}(x_i-x_j)f_{x,i,j}\right\rangle_t
\label{eq:Pxfinal}
\end{equation}
because of the relation that $f_{x,i,j}=-f_{x,j,i}$.
Equation (\ref{eq:Pxfinal}) is different from that of \citet{Kataoka2013} because we consider the tensile stress only in the $x$-axis direction.

We take an average of the tensile stress $P_x$ for every 10,000 time-steps, at least.
In some simulations, the tensile stress fluctuates, and thus we take a longer time average to smooth it (see Section \ref{subsec:fiducial} for details).
One time-step in our simulation is $0.7t_\mathrm{c}=1.9\times10^{-11}$ s, and therefore 10,000 time-steps correspond to $1.9\times10^{-7}$ s.

\subsection{Overview of Our Simulations} \label{subsec:overview}

\begin{figure*}
\plotone{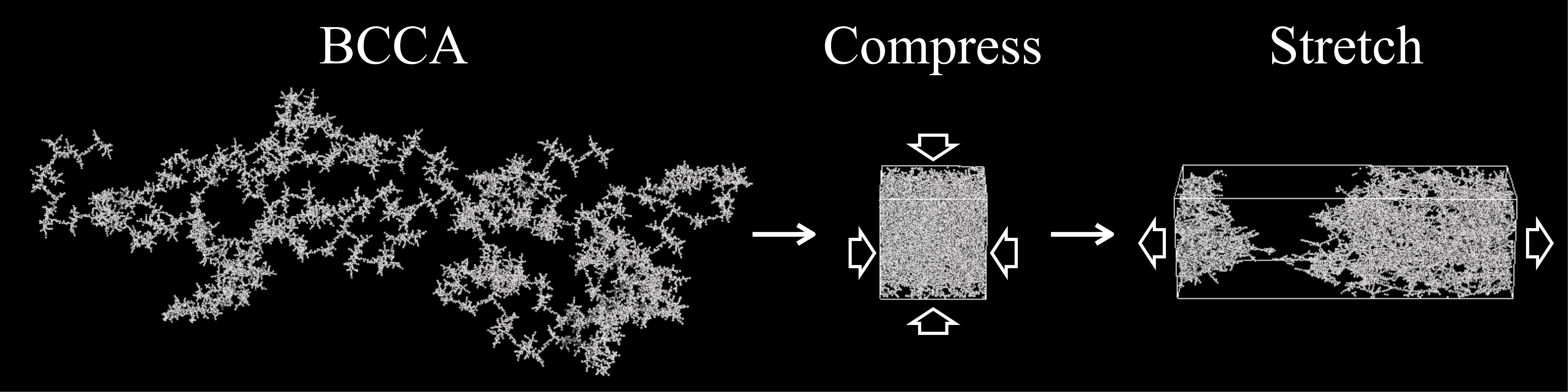}
\caption{Overview of our simulations.
Each picture shows a BCCA (left), a compressed aggregate (center), and a stretched aggregate (right).
Each aggregate contains 16384 ice monomers whose radius is 0.1 $\mathrm{\mu m}$.
In the center and right panels, the box with white lines shows the calculation box with periodic boundaries.}
\label{fig:overview}
\end{figure*}

The overview of our numerical simulations is as follows.
First, we randomly create a BCCA to change the initial condition.
Next, we compress it statically and isotropically \citep{Kataoka2013}.
The compression of a BCCA corresponds to the formation of a planetesimal.
It is necessary for the BCCA to be attached to all boundaries so that we can stretch it.
Then, we stop compression and define this volume filling factor as the initial one $\phi_\mathrm{init}$.
Finally, we stretch it statically and one-dimensionally.
Figure \ref{fig:overview} shows the overview of our simulations.

\section{Results} \label{sec:results}

We perform 10 simulations with different initial dust aggregates for every parameter set.
First, we perform fiducial runs to investigate what occurs in our stretching simulations in Section \ref{subsec:fiducial}.
Then, we show that the results do not depend on any numerical parameters, such as the number of particles $N$, boundary velocity $v_\mathrm{b}$, and the damping coefficient $k_\mathrm{n}$ in Section \ref{subsec:numparamdepend}.
Finally, in Section \ref{subsec:phyparamdepend}, we investigate the dependence on physical parameters: the initial volume filling factor $\phi_\mathrm{init}$, monomer radius $r_0$, surface energy $\gamma$, and the critical rolling displacement $\xi_\mathrm{crit}$.

\subsection{Fiducial Run} \label{subsec:fiducial}

\begin{figure}
\plotone{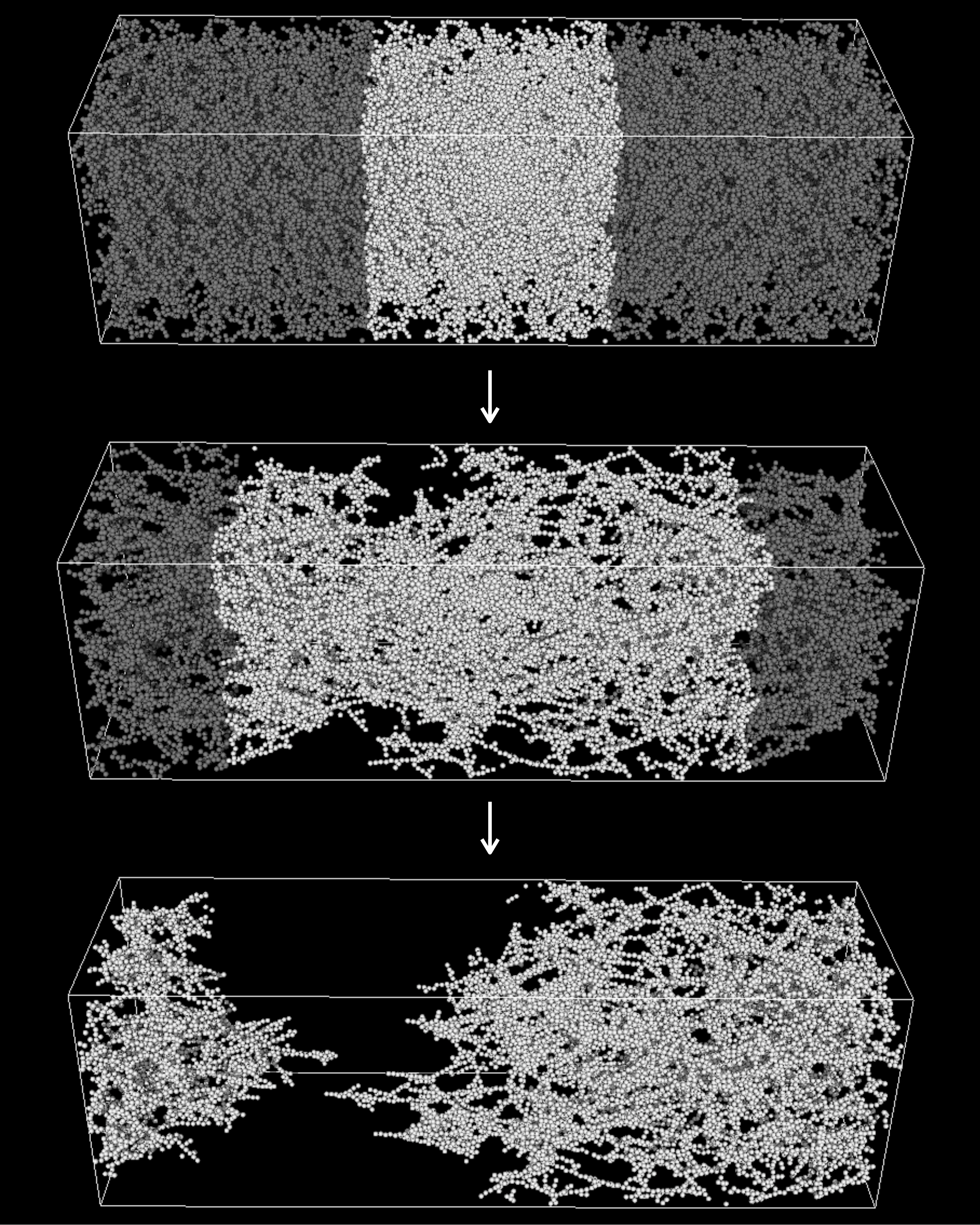}
\caption{Snapshots of a fiducial run when $N=16384$, $v_\mathrm{b}=10\mathrm{\ cm\ s^{-1}}$, $k_\mathrm{n}=0.01$, $\phi_\mathrm{init}=0.1$, $r_0=0.1\mathrm{\ \mu m}$, $\gamma=100\mathrm{\ mJ\ m^{-2}}$, and $\xi_\mathrm{crit}=8$ \AA.
The calculation times are $t=0\mathrm{\ s}$ (top), $t=2.49\times10^5t_\mathrm{step}\sim4.7\times10^{-6}\mathrm{\ s}$ (center), and $t=4.98\times10^5t_\mathrm{step}\sim9.5\times10^{-6}\mathrm{\ s}$ (bottom), where $t_\mathrm{step}=0.7t_\mathrm{c}=1.9\times10^{-11}\mathrm{\ s}$ represents one time-step.
Each particle represents a 0.1 $\mathrm{\mu m}$-radius ice monomer.
The light gray monomers are in the calculation box with periodic boundaries, while the dark gray monomers are in the neighbor boxes.
The box with white lines shows the final state of the calculation box.
We omit the boxes in front, behind, top, and bottom of the calculation box for simplicity.}
\label{fig:snap}
\end{figure}

We measure the tensile stress of 10 runs for the fiducial parameter set.
The fiducial values are $N=16384$, $v_\mathrm{b}=10\mathrm{\ cm\ s^{-1}}$, $k_\mathrm{n}=0.01$, $\phi_\mathrm{init}=0.1$, $r_0=0.1\mathrm{\ \mu m}$, $\gamma=100\mathrm{\ mJ\ m^{-2}}$, and $\xi_\mathrm{crit}=8$ \AA.
Figure \ref{fig:snap} shows three snapshots of a fiducial run.
Each particle represents a 0.1 $\mathrm{\mu m}$-radius ice monomer.
The light gray monomers are in the calculation box with periodic boundaries, while the dark gray monomers are in the neighbor boxes.
The box with white lines shows the final state of the calculation box.
By stretching the dust aggregate, the chain-like structure appears.

\begin{figure*}
\plottwo{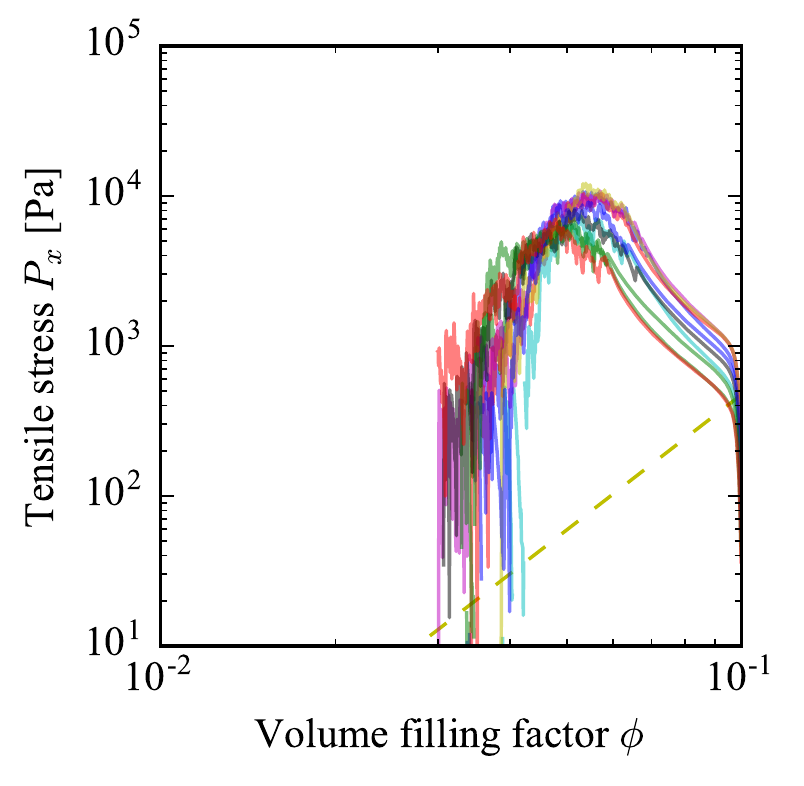}{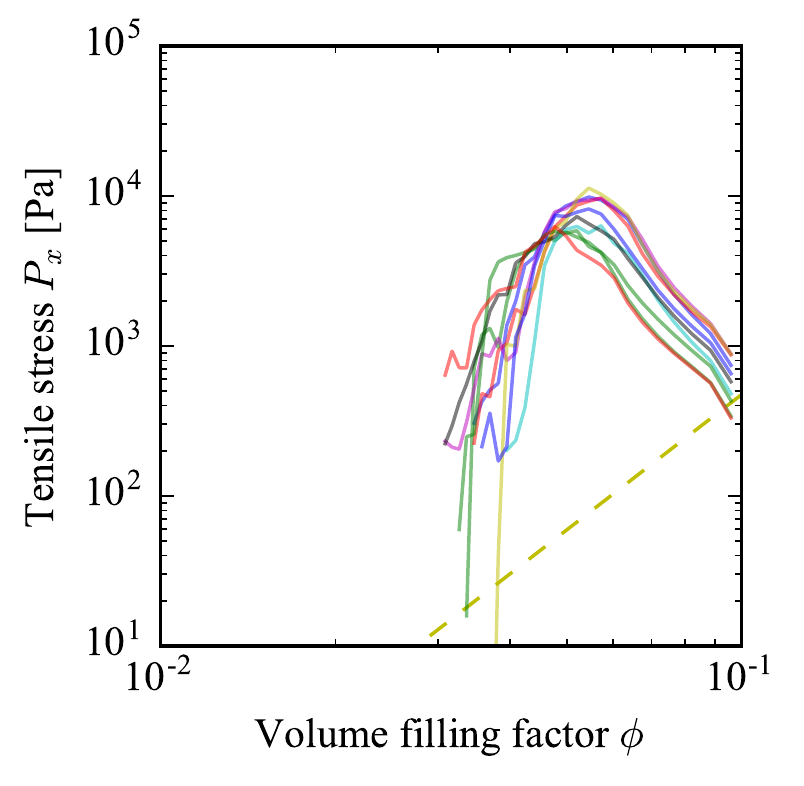}
\caption{Tensile stress $P_x$ of 10 fiducial runs averaged for every 10,000 time-steps (left) and 200,000 time-steps (right) when $N=16384$, $v_\mathrm{b}=10\mathrm{\ cm\ s^{-1}}$, $k_\mathrm{n}=0.01$, $\phi_\mathrm{init}=0.1$, $r_0=0.1\mathrm{\ \mu m}$, $\gamma=100\mathrm{\ mJ\ m^{-2}}$, and $\xi_\mathrm{crit}=8$ \AA.
The yellow dashed lines show compressive strength (Equation (\ref{eq:Pcomp})) investigated by \citet{Kataoka2013}.}
\label{fig:fiducial}
\end{figure*}

Figure \ref{fig:fiducial} shows the time evolution of tensile stress of 10 fiducial runs averaged for every 10,000 time-steps (left) and 200,000 time-steps (right).
The volume filling factor at each time-step is calculated as
\begin{equation}
\phi = \frac{(4/3)\pi r_0^3N}{V}
\end{equation}
and the tensile stress is calculated according to Equation (\ref{eq:Pxfinal}).
We choose the number of time-steps when the rate of change of the volume filling factor does not exceed 10\%.
As tensile displacement increases, the volume filling factor $\phi$ decreases and the tensile stress $P_x$ increases.
The maximum value of tensile stress is called the tensile strength.
To calculate the tensile strength for every parameter set, we find 10 maximum values of the obtained tensile stress and take an average of them.

\subsection{Numerical Parameter Dependence} \label{subsec:numparamdepend}

\begin{figure*}
\plotone{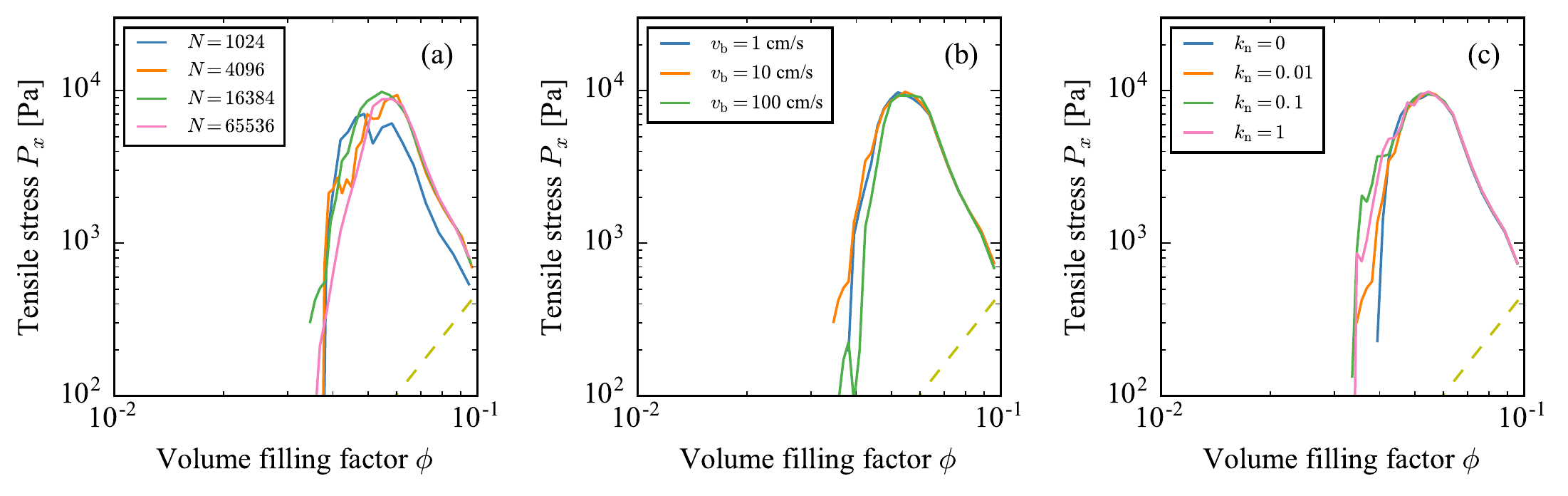}
\caption{Tensile stress $P_x$ with different numbers of particles $N$ (left), different boundary velocities $v_\mathrm{b}$ (center), and different damping coefficients $k_\mathrm{n}$ (right).
The fiducial values are $N=16384$, $v_\mathrm{b}=10\mathrm{\ cm\ s^{-1}}$, $k_\mathrm{n}=0.01$, $\phi_\mathrm{init}=0.1$, $r_0=0.1\mathrm{\ \mu m}$, $\gamma=100\mathrm{\ mJ\ m^{-2}}$, and $\xi_\mathrm{crit}=8$ \AA.}
\label{fig:numparam}
\end{figure*}

To investigate the dependence on the number of particles $N$, we plot tensile stress when $N=2^{10}=1024$, $N=2^{12}=4096$, $N=2^{14}=16384$, and $N=2^{16}=65536$ in Figure \ref{fig:numparam}(a).
Changing $N$ corresponds to changing the size of the calculation box.
Obviously, the tensile strength has no dependence on $N$.
Because of the smoothness of the tensile stress plot and calculation costs, we set $N=16384$ as the fiducial value.

Figure \ref{fig:numparam}(b) shows tensile stress when boundary velocity $v_\mathrm{b}=1\mathrm{\ cm\ s^{-1}}$, $10\mathrm{\ cm\ s^{-1}}$, and $100\mathrm{\ cm\ s^{-1}}$.
All boundary velocities are less than the effective sound speed of dust aggregates $c_\mathrm{s,eff}$ (Equation (\ref{eq:sound2})).
There is no difference among the three boundary velocities.
Therefore, we can conclude that dust aggregates are stretched statically.
We set $v_\mathrm{b}=10\mathrm{\ cm\ s^{-1}}$ as the fiducial value considering sampling rates of tensile stress and calculation costs.

Tensile stress with various damping coefficients is plotted in Figure \ref{fig:numparam}(c).
No damping force corresponds to $k_\mathrm{n}=0$.
We change the strength of damping force from weak damping ($k_\mathrm{n}=0.01$) to strong damping ($k_\mathrm{n}=1$).
Undoubtedly, there is no dependence on the damping force in this range.
We use $k_\mathrm{n}=0.01$ for all the other simulations.

\subsection{Physical Parameter Dependence} \label{subsec:phyparamdepend}

\begin{figure}
\plotone{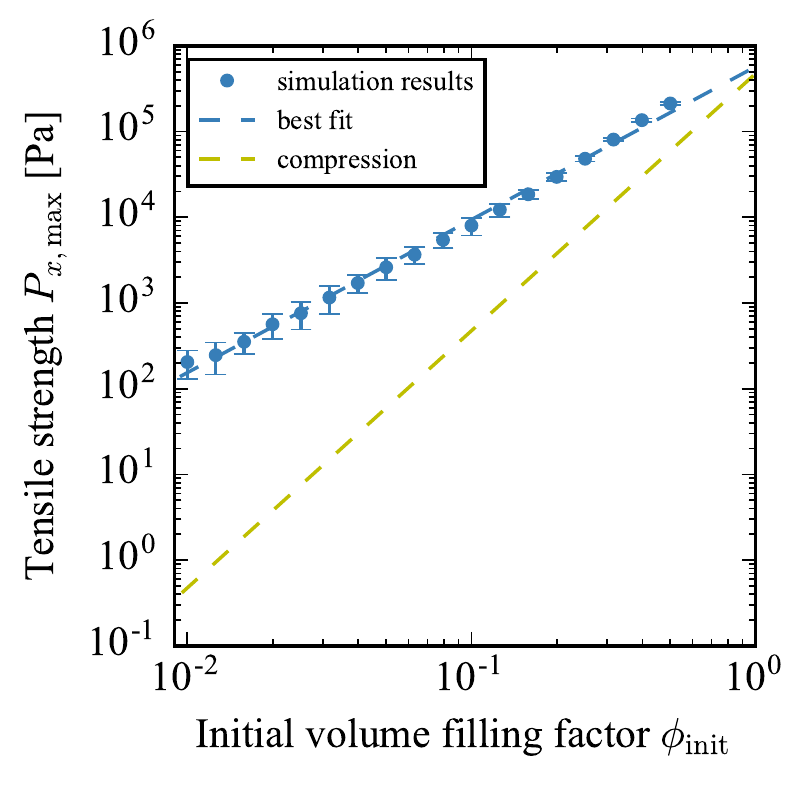}
\caption{Tensile strength $P_{x,\mathrm{max}}$ as a function of initial volume filling factor $\phi_\mathrm{init}$ when $N=16384$, $v_\mathrm{b}=10\mathrm{\ cm\ s^{-1}}$, $k_\mathrm{n}=0.01$, $r_0=0.1\mathrm{\ \mu m}$, $\gamma=100\mathrm{\ mJ\ m^{-2}}$, and $\xi_\mathrm{crit}=8$ \AA.
The blue and yellow dashed lines show the best fit for the tensile strength (Equation (\ref{eq:fitphi})) and the compressive strength (Equation (\ref{eq:Pcomp})), respectively.
The error bar corresponds to the standard deviation of 10 runs.}
\label{fig:tensile}
\end{figure}

We measure the tensile strength of dust aggregates which have various initial volume filling factors as shown in Figure \ref{fig:tensile}.
The calculated tensile strength is proportional to $\phi_\mathrm{init}^{1.8}$ from the fitting.
The tensile strength $P_{x,\mathrm{max}}$ can be described with the initial volume filling factor $\phi_\mathrm{init}$ as
\begin{equation}
P_{x,\mathrm{max}} = P_0\phi_\mathrm{init}^{1.8},
\label{eq:fitphi}
\end{equation}
where $P_0\sim6\times10^5$ Pa in this case.
The analytical interpretation of Equation (\ref{eq:fitphi}) is discussed in Section \ref{subsec:semianamodel}.

\begin{figure}
\plotone{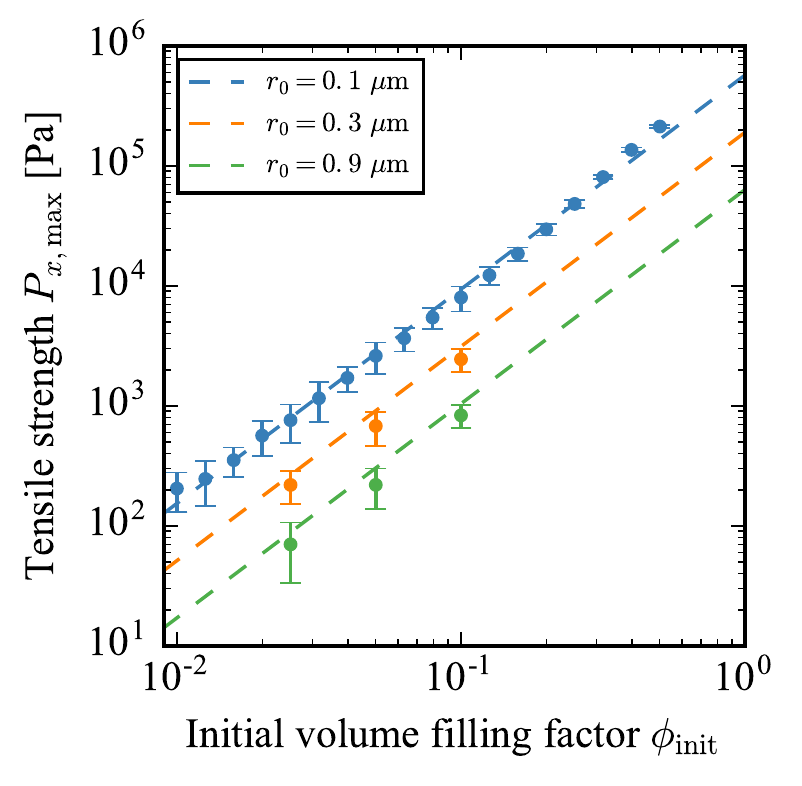}
\caption{Tensile strength $P_{x,\mathrm{max}}$ as a function of initial volume filling factor $\phi_\mathrm{init}$ when $N=16384$, $k_\mathrm{n}=0.01$, $\gamma=100\mathrm{\ mJ\ m^{-2}}$, and $\xi_\mathrm{crit}=8$ \AA.
The monomer radii are $r_0=0.1\mathrm{\ \mu m}$ (blue), $r_0=0.3\mathrm{\ \mu m}$ (orange), and $r_0=0.9\mathrm{\ \mu m}$ (green).
The error bar corresponds to the standard deviation of 10 runs.
The dashed lines represent Equation (\ref{eq:Pmaxequation}).}
\label{fig:tensileradius}
\end{figure}

To investigate the dependence of tensile strength on the monomer radius, we perform simulations in the case of ice monomers whose radii are 0.3 $\mathrm{\mu m}$ and 0.9 $\mathrm{\mu m}$.
Figure \ref{fig:tensileradius} shows the summary of the monomer radius dependence.
The plotted dashed lines are based on Equation (\ref{eq:Pmaxequation}), which is an analytical expression of tensile strength (see Section \ref{subsec:semianamodel}).
It is confirmed that the tensile strength is in inverse proportion to the monomer radius.

\begin{figure}
\plotone{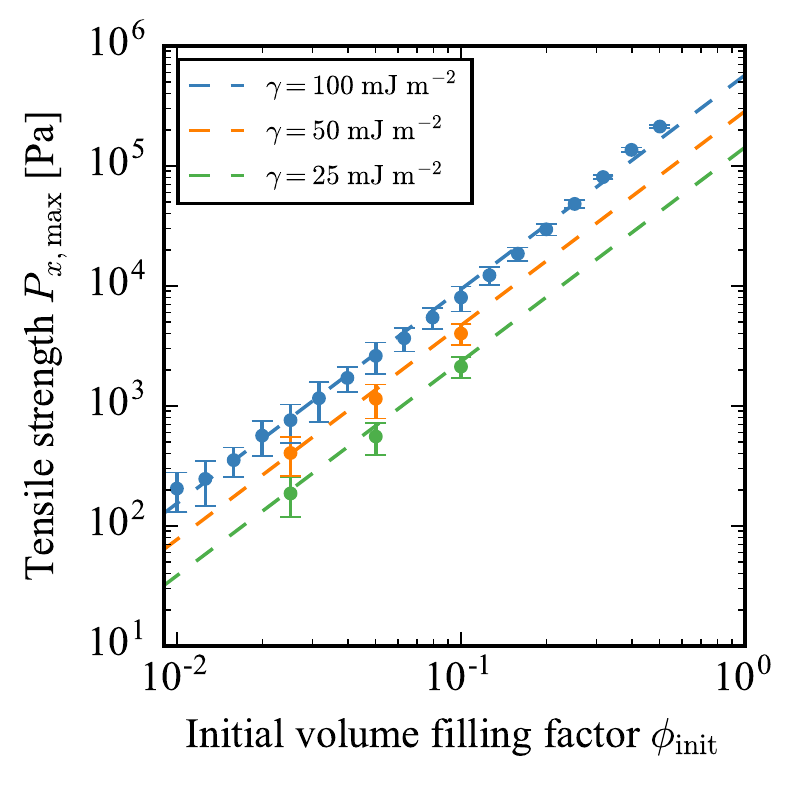}
\caption{Tensile strength $P_{x,\mathrm{max}}$ as a function of initial volume filling factor $\phi_\mathrm{init}$ when $N=16384$, $k_\mathrm{n}=0.01$, $r_0=0.1\mathrm{\ \mu m}$, and $\xi_\mathrm{crit}=8$ \AA.
The values of surface energy are $\gamma=100\mathrm{\ mJ\ m^{-2}}$ (blue), $\gamma=50\mathrm{\ mJ\ m^{-2}}$ (orange), and $\gamma=25\mathrm{\ mJ\ m^{-2}}$ (green).
The error bar corresponds to the standard deviation of 10 runs.
The dashed lines represent Equation (\ref{eq:Pmaxequation}).}
\label{fig:tensilegamma}
\end{figure}

To clarify the surface energy dependence, we calculate the tensile strength when $\gamma=50\mathrm{\ mJ\ m^{-2}}$ and $25\mathrm{\ mJ\ m^{-2}}$ and plot it in Figure \ref{fig:tensilegamma}.
Other parameters are the same as the fiducial values.
The dashed lines represent Equation (\ref{eq:Pmaxequation}) (see Section \ref{subsec:semianamodel}).
Obviously, the tensile strength is in proportion to the surface energy.

\begin{figure}
\plotone{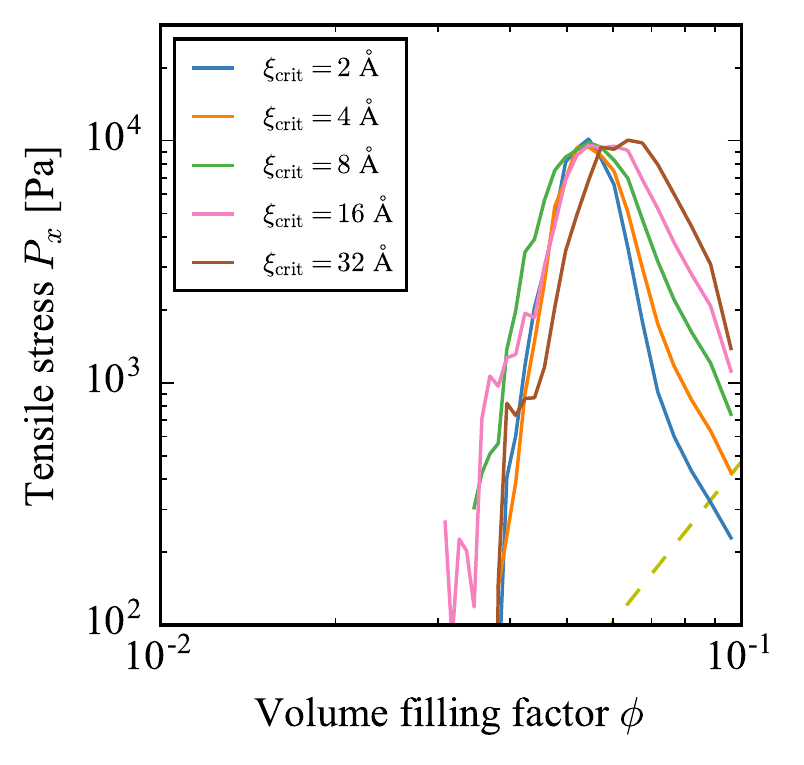}
\caption{Tensile stress $P_x$ with different critical rolling displacements $\xi_\mathrm{crit}$ when $N=16384$, $v_\mathrm{b}=10\mathrm{\ cm\ s^{-1}}$, $k_\mathrm{n}=0.01$, $\phi_\mathrm{init}=0.1$, $r_0=0.1\mathrm{\ \mu m}$, and $\gamma=100\mathrm{\ mJ\ m^{-2}}$.
The critical rolling displacements are $\xi_\mathrm{crit}=2$ \AA (blue), $\xi_\mathrm{crit}=4$ \AA (orange), $\xi_\mathrm{crit}=8$ \AA (green), $\xi_\mathrm{crit}=16$ \AA (magenta), and $\xi_\mathrm{crit}=32$ \AA (brown).}
\label{fig:xi}
\end{figure}

Finally, we investigate the dependence of tensile stress on the critical rolling displacement $\xi_\mathrm{crit}$ in Figure \ref{fig:xi}.
The critical rolling displacement is changed from $\xi_\mathrm{crit}=2\textrm{\ \AA}$ \\citep[e.g.,][]{Dominik1997} to $\xi_\mathrm{crit}=32\textrm{\ \AA}$ \citep[e.g.,][]{Heim1999}.
Tensile stress has a marginal dependence on $\xi_\mathrm{crit}$ because the main mechanism of displacement is rolling (see Section \ref{subsec:semianamodel}).
On the other hand, tensile strength, which is the maximum value of tensile stress, has no difference.
We can conclude that the tensile strength is almost the same even if the critical rolling displacement has uncertainty.
Therefore, we fix $\xi_\mathrm{crit}=8\textrm{\ \AA}$ in our simulations.

\section{Discussions} \label{sec:discuss}

Now, we discuss the obtained physical parameter dependence of the tensile strength of dust aggregates (Section \ref{subsec:phyparamdepend}) and apply our results to previous studies of experiments, numerical simulations, and comet 67P.
In Section \ref{subsec:semianamodel}, we find an analytical expression of the tensile strength using material parameters: the initial volume filling factor, monomer radius, and the surface energy.
Then, we compare our results with previous experiments and numerical simulations about silicate dust aggregates \citep{Blum2004,Blum2006,Seizinger2013,Gundlach2018} in Section \ref{subsec:compare}.
Finally, we apply our interpretation to comet 67P in Section \ref{subsec:apply}.

\subsection{Semi-Analytical Model of Tensile Strength} \label{subsec:semianamodel}

The relationship between $P_{x,\mathrm{max}}$ and $\phi_\mathrm{init}$ can be derived by considering the maximum force needed to separate two sticking monomers $F_\mathrm{c}$ and the radius of a dust aggregate $r_\mathrm{agg}$.
When the tensile stress has a maximum value, the force $F_\mathrm{c}$ is applied on a connection between two monomers of a dust aggregate.
This means that
\begin{equation}
P_{x,\mathrm{max}} \propto \frac{F_\mathrm{c}}{r_\mathrm{agg}^2}.
\label{eq:Pmax}
\end{equation}
The radius of a dust aggregate is given as
\begin{equation}
r_\mathrm{agg} \propto N_\mathrm{agg}^{1/D}r_0,
\label{eq:aggfractal}
\end{equation}
where $D$ and $N_\mathrm{agg}$ are the fractal dimension and the number of monomers of a dust aggregate, respectively.
The initial volume filling factor is described as
\begin{equation}
\phi_\mathrm{init} = N_\mathrm{agg}\left(\frac{r_0}{r_\mathrm{agg}}\right)^3,
\end{equation}
and then, the radius of a dust aggregate is obtained as
\begin{equation}
r_\mathrm{agg} \propto r_0\phi_\mathrm{init}^{-1/(3-D)}.
\end{equation}
From Equation (\ref{eq:Pmax}), the tensile strength can be written as
\begin{equation}
P_{x,\mathrm{max}} \sim C \frac{F_\mathrm{c}}{r_0^2}\phi_\mathrm{init}^{2/(3-D)},
\end{equation}
where $C$ is a constant.
The fractal dimension $D$ of BCCAs is $\sim1.9$ \citep{Mukai1992,Okuzumi2009dustagg}.

Using the fitting result of Equations (\ref{eq:Fc}) and (\ref{eq:fitphi}), we obtain
\begin{eqnarray}
P_{x,\mathrm{max}} &\sim&6\times10^5\left(\frac{\gamma}{100\mathrm{\ mJ\ m^{-2}}}\right)\nonumber\\
&&\times\left(\frac{r_0}{0.1\mathrm{\ \mu m}}\right)^{-1}\phi_\mathrm{init}^{1.8}\mathrm{\ Pa}.
\label{eq:Pmaxequation}
\end{eqnarray}
In the case of ice monomer whose radius is 0.1 $\mathrm{\mu m}$, we find $C=0.12\pm0.01$.

\begin{figure}
\plotone{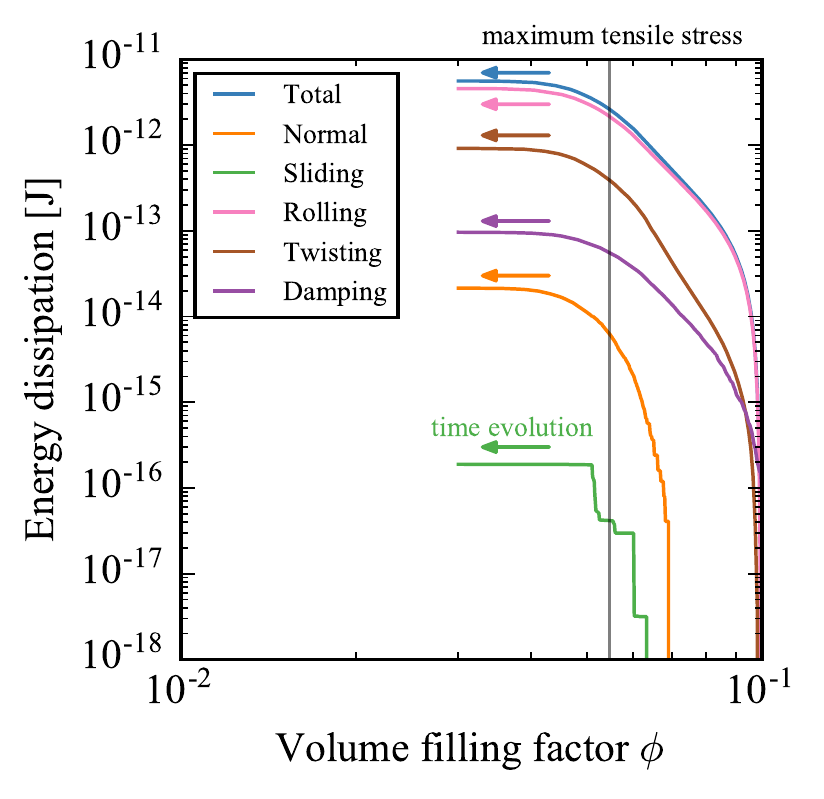}
\caption{Energy dissipations of a fiducial run when $N=16384$, $v_\mathrm{b}=10\mathrm{\ cm\ s^{-1}}$, $k_\mathrm{n}=0.01$, $\phi_\mathrm{init}=0.1$, $r_0=0.1\mathrm{\ \mu m}$, $\gamma=100\mathrm{\ mJ\ m^{-2}}$, and $\xi_\mathrm{crit}=8$ \AA.
The energy dissipation mechanisms are the normal (orange), sliding (green), rolling (magenta), twisting (brown), damping (purple), and the total of all energy dissipations (blue).
The vertical gray line represents the volume filling factor when tensile stress has a maximum value.}
\label{fig:energy}
\end{figure}

We confirm Equation (\ref{eq:Pmaxequation}) from the perspective of energy dissipation.
All energy dissipations, which are caused by the normal, sliding, rolling, twisting, and damping force in the normal direction, are plotted in Figure \ref{fig:energy}.
The curves in Figure \ref{fig:energy} run time-wise from right to left and arise during the stretching of a dust aggregate.
The main energy dissipation mechanism is the rolling, which corresponds to the $\xi_\mathrm{crit}$-dependence of tensile stress (Section \ref{subsec:phyparamdepend}).
The energy dissipation by the normal arises when the tensile stress has a maximum value.
This energy dissipation is caused by connection breaking between two contacting monomers.
For this reason, tensile strength is determined by the connection breaking, i.e. $F_\mathrm{c}$.

\begin{figure}
\plotone{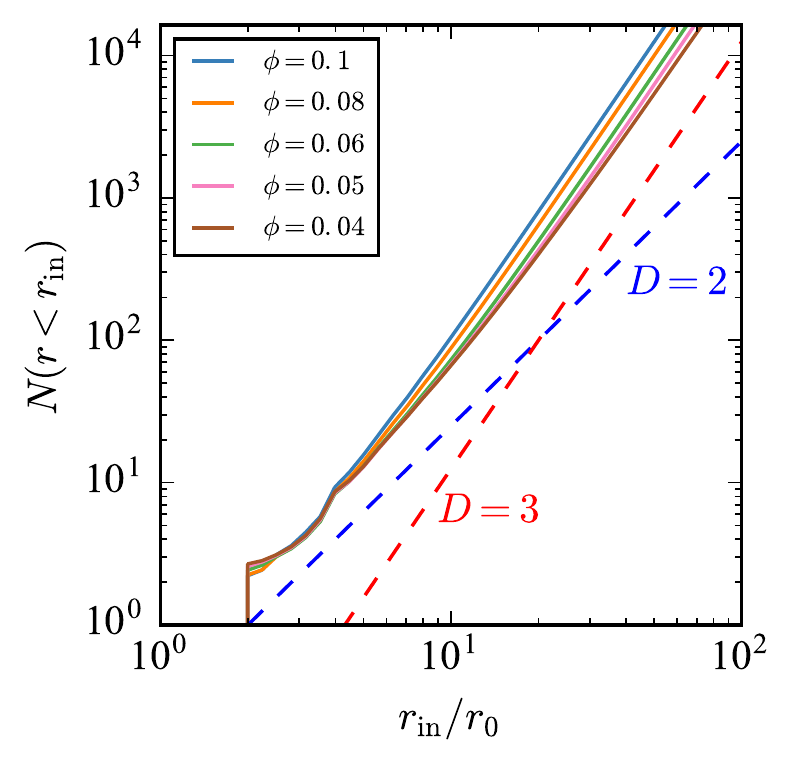}
\caption{Number of monomers inside the radius $r_\mathrm{in}$ as a function of $r_\mathrm{in}/r_0$ when $N=16384$, $v_\mathrm{b}=10\mathrm{\ cm\ s^{-1}}$, $k_\mathrm{n}=0.01$, $\phi_\mathrm{init}=0.1$, $r_0=0.1\mathrm{\ \mu m}$, $\gamma=100\mathrm{\ mJ\ m^{-2}}$, and $\xi_\mathrm{crit}=8$ \AA.
Five snapshots are represented by blue ($\phi=\phi_\mathrm{init}=0.1$), orange ($\phi=0.08$), green ($\phi=0.06$), magenta ($\phi=0.05$), and brown ($\phi=0.04$) lines.
The blue and red dashed lines show the relationship when $D=2$ and $D=3$, respectively.}
\label{fig:fractal}
\end{figure}

To confirm that $D\sim1.9$ on a small scale of a dust aggregate in our simulations, we calculate the number of monomers inside the radius $r_\mathrm{in}$ for five snapshots of a fiducial run and plot it in Figure \ref{fig:fractal}.
We take the snapshots during the continuous strain of the dust aggregate.
The parameters of the fiducial run are $N=16384$, $v_\mathrm{b}=10\mathrm{\ cm\ s^{-1}}$, $k_\mathrm{n}=0.01$, $\phi_\mathrm{init}=0.1$, $r_0=0.1\mathrm{\ \mu m}$, $\gamma=100\mathrm{\ mJ\ m^{-2}}$, and $\xi_\mathrm{crit}=8$ \AA.
The method to count the number of monomers $N(r<r_\mathrm{in})$ is as follows.
At first, we set a monomer in the calculation box as the center and count $N(r<r_\mathrm{in})$ including monomers outside the periodic boundaries.
Next, we take an average of $N(r<r_\mathrm{in})$ for all monomers in the calculation box.

Also, we plot $N(r<r_\mathrm{in})$ as a function of $r_\mathrm{in}/r_0$ when $D=2$ and $D=3$ in Figure \ref{fig:fractal}.
This relationship is derived by Equation (\ref{eq:aggfractal}) as
\begin{equation}
N(r<r_\mathrm{in}) \propto \left(\frac{r_\mathrm{in}}{r_0}\right)^D.
\end{equation}
On a small scale that $N(r<r_\mathrm{in})\lesssim20$, all snapshot results mostly correspond to the relationship when $D=2$.
When the scale becomes large enough, dust aggregates have a fractal dimension of three.

\subsection{Comparison with Previous Studies} \label{subsec:compare}

\begin{figure*}
\plottwo{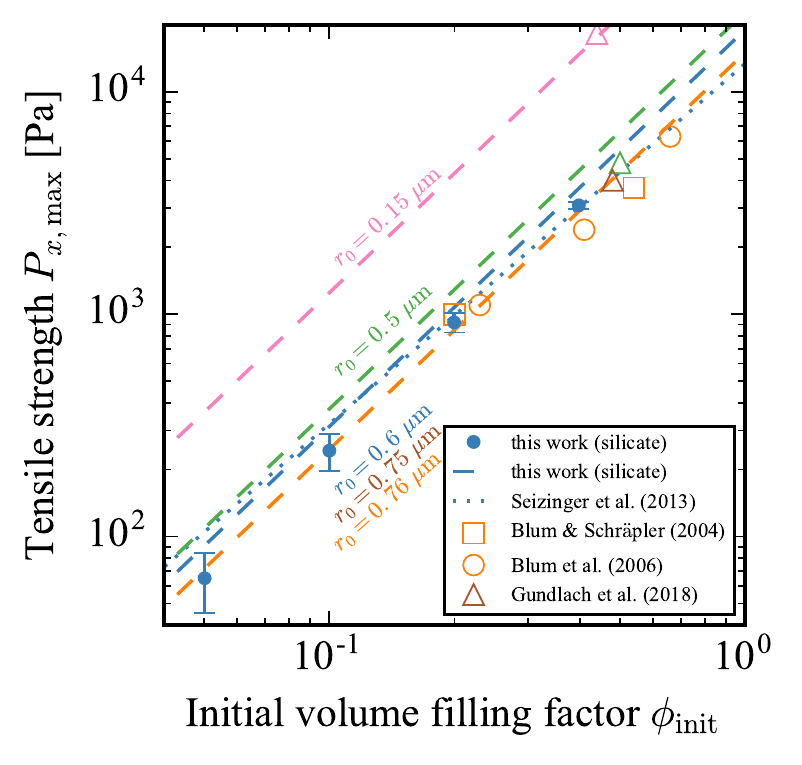}{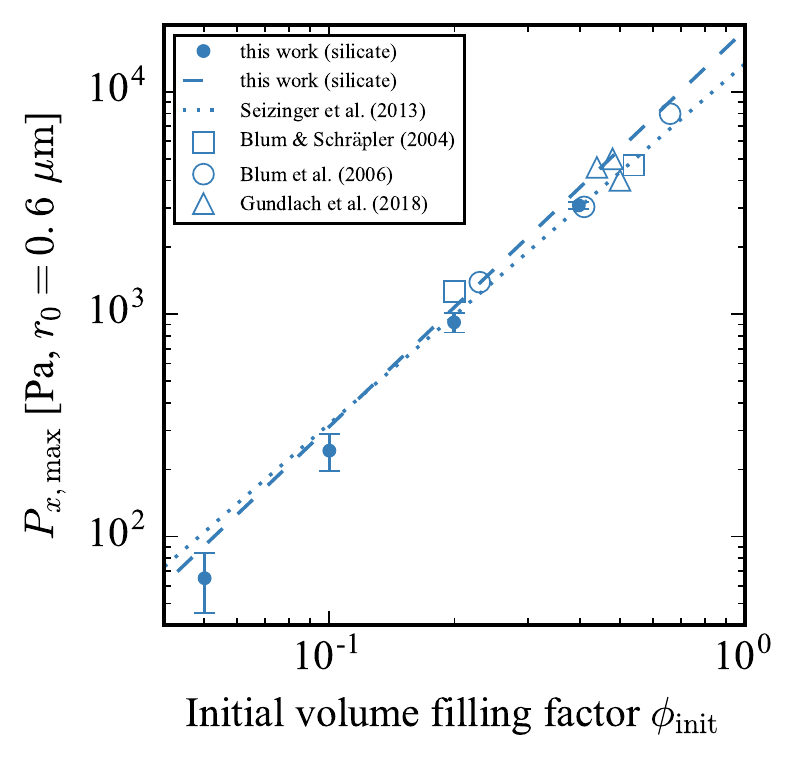}
\caption{Tensile strength $P_{x,\mathrm{max}}$ of silicate dust aggregates of this work and previous studies, which contain various initial volume filling factors $\phi_\mathrm{init}$ and monomer radii $r_0$ (left), and that scaled to monomer radius of 0.6 $\mathrm{\mu m}$ (right).
The filled circles show our simulation results when $N=16384$ and $k_\mathrm{n}=0.01$.
Other material parameters of silicate are listed in Table \ref{tab:param}.
The error bar corresponds to the standard deviation of 10 runs.
The dashed lines represent Equation (\ref{eq:Pmaxequation}), while the dotted line represents Equation (2) of \citet{Seizinger2013}.
The experimental results are denoted by open squares \citep{Blum2004}, open circles \citep{Blum2006}, and open triangles \citep{Gundlach2018}.
The monomer radii are $0.15\mathrm{\ \mu m}$ (magenta), $0.5\mathrm{\ \mu m}$ (green), $0.6\mathrm{\ \mu m}$ (blue), $0.75\mathrm{\ \mu m}$ (brown), and $0.76\mathrm{\ \mu m}$ (orange).}
\label{fig:tensilesilicate}
\end{figure*}

To compare our results with previous experiments and numerical simulations, we perform simulations with the parameters of silicate listed in Table \ref{tab:param} and summarize the results in Figure \ref{fig:tensilesilicate}.
Both results by experiments \citep{Blum2004,Blum2006,Gundlach2018} and simulations \citep[this work;][]{Seizinger2013} correspond very well.
This means that there is little influence of artificial adhesion force introduced by \citet{Seizinger2013} and small aggregates whose sizes are from $\mathrm{\mu m}$ to mm.
The right panel of Figure \ref{fig:tensilesilicate} is derived by Equation (\ref{eq:Pmaxequation}), i.e., $P_{x,\mathrm{max}}\propto r_0^{-1}$.

\begin{figure}
\plotone{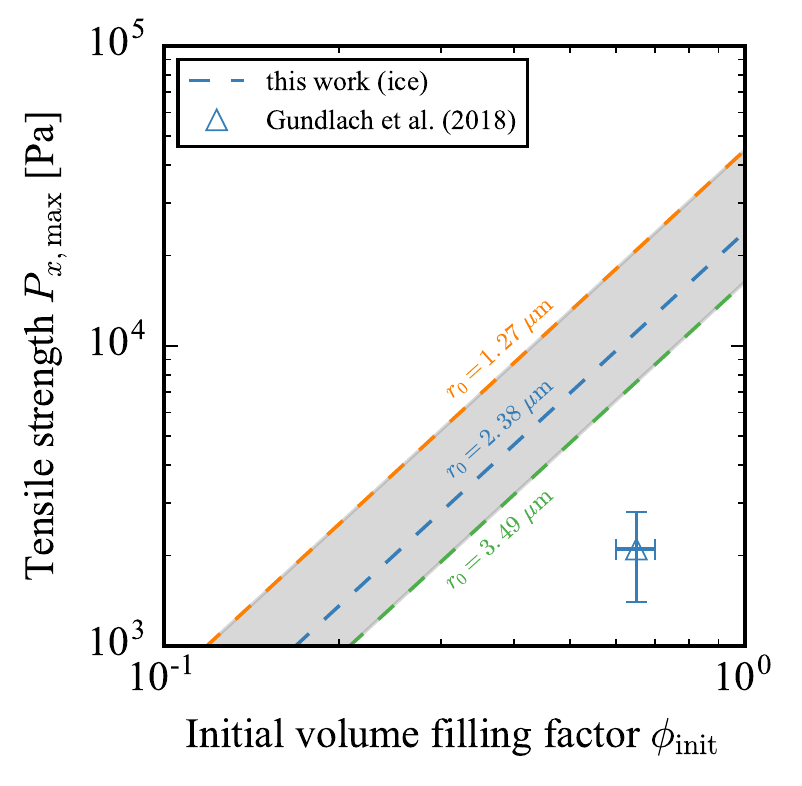}
\caption{Tensile strength $P_{x,\mathrm{max}}$ of ice dust aggregates of this work and the previous study.
The dashed lines represent Equation (\ref{eq:Pmaxequation}).
The experimental result when $r_0=2.38\pm1.11\mathrm{\ \mu m}$ is denoted by the open triangle with error bars \citep{Gundlach2018}.
The monomer radii are $1.27\mathrm{\ \mu m}$ (orange), $2.38\mathrm{\ \mu m}$ (blue), and $3.49\mathrm{\ \mu m}$ (green).}
\label{fig:tensileice}
\end{figure}

\citet{Gundlach2018} measured the tensile strength of ice aggregates with the monomer radius of $2.38\pm1.11\mathrm{\ \mu m}$, which is shown in Figure \ref{fig:tensileice}.
We do not know which monomer radius determines the tensile strength of dust aggregates when the monomer radius has a size distribution.
Therefore, we plot dashed lines derived by Equation (\ref{eq:Pmaxequation}) when $r_0=1.27\mathrm{\ \mu m}$, $r_0=2.38\mathrm{\ \mu m}$, and $r_0=3.49\mathrm{\ \mu m}$.
The experimental value is lower than the theoretical lines.
From their low value of the tensile strength, \citet{Gundlach2018} inferred that the specific surface energy of ice, $\gamma_\mathrm{ice}$, has a very value of $=0.02\mathrm{\ J\ m^{-2}}$ at low temperatures ($\lesssim150$ K).
However, \citet{Gundlach2015} also estimated as $\gamma_\mathrm{ice}=0.19\mathrm{\ J\ m^{-2}}$ from the constant sticking threshold velocity of $\sim10\mathrm{\ m\ s^{-1}}$ for $T<210$ K in their ice impact experiments.
The reason for low tensile strength in \citet{Gundlach2018} is thus unknown.
We also confirm that the experimental value is higher than the compressive strength theoretically expected by \citet{Kataoka2013} (Equation (\ref{eq:Pcomp})).

\subsection{Application to Comet 67P} \label{subsec:apply}

We can estimate the monomer radius of comet 67P using Equation (\ref{eq:Pmaxequation}) as follows.
From the exploration of 67P, it was found that the micro-porosity of 67P is 0.75--0.85 \citep{Kofman2015}, while the surface porosity is 0.87 \citep{Fornasier2015}.
In other words, the volume filling factor of 67P is about 0.13--0.25.
The averaged nucleus bulk density of 67P was estimated to be 0.533 $\mathrm{g\ cm^{-3}}$ \citep{Patzold2016}.
In consideration of the high dust-to-water mass ratio of 67P \citep[e.g.,][]{Fulle2016}, its main material is not H$_2$O ice, but silicate.
Assuming that the bulk density is 0.533 $\mathrm{g\ cm^{-3}}$ and the material density is 2.65 $\mathrm{g\ cm^{-3}}$ (Table \ref{tab:param}), we can estimate that the volume filling factor of 67P is about 0.20, which is consistent with the values of 0.13--0.25.
Substituting $P_{x,\mathrm{max}}=3$--200 Pa (see Section \ref{sec:intro}), $\gamma=20\mathrm{\ mJ\ m^{-2}}$, and $\phi_\mathrm{init}=0.20$ in Equation (\ref{eq:summaryeq}), we find that the monomer radius of 67P has to be 3.3--220 $\mathrm{\mu m}$.
To explain the low tensile strength of 67P using only our model, we have to consider a larger radius than that of interstellar materials.

Another idea to decrease the tensile strength is assuming an aggregate of aggregates \citep[e.g.,][]{Blum2014,Blum2017} instead of a simple aggregate of monomers as discussed in this paper.
The aggregate-of-aggregate model may explain the low strength with small monomers, but this is beyond the scope of this paper.

\section{Conclusions} \label{sec:conclusions}

We investigated the tensile strength of porous dust aggregates whose initial volume filling factors $\phi_\mathrm{init}$ are from $10^{-2}$ to 0.5.
We performed three-dimensional numerical simulations with periodic boundary condition to measure the tensile stress of dust aggregates.
The monomer interaction model is based on \citet{Dominik1997} and \citet{Wada2007}.
The initial dust aggregates are statically and isotropically compressed BCCAs investigated by \citet{Kataoka2013}.
At boundaries of the calculation box, we set moving periodic boundaries in the $x$-axis direction and fixed periodic boundaries in the $y$- and $z$-axis directions.
The calculation method of the tensile stress is the same way as molecular dynamics.
In our simulations, we created a BCCA at first, compressed it three-dimensionally, and then stretched it one-dimensionally.
For every parameter set, we conducted 10 stretching-simulations with different initial dust aggregates, found 10 maximum values of the obtained tensile stress, and took an average of them, which is called the tensile strength.
Our main findings of the tensile strength of porous dust aggregates are as follows.

\begin{itemize}
\item As a result of numerical simulations, we found that the tensile strength $P_{x,\mathrm{max}}$ can be written as
\begin{eqnarray}
P_{x,\mathrm{max}} &\sim& 0.12\frac{F_\mathrm{c}}{r_0^2}\phi_\mathrm{init}^{1.8}\nonumber\\
&\sim&6\times10^5\left(\frac{\gamma}{100\mathrm{\ mJ\ m^{-2}}}\right)\nonumber\\
&&\times\left(\frac{r_0}{0.1\mathrm{\ \mu m}}\right)^{-1}\phi_\mathrm{init}^{1.8}\mathrm{\ Pa},
\label{eq:summaryeq}
\end{eqnarray}
where $F_\mathrm{c}$ is the maximum force needed to separate two sticking monomers, $\phi_\mathrm{init}$ is the initial volume filling factor, $r_0$ is the monomer radius, and $\gamma$ is the surface energy.

\item We analytically confirmed the dependence in Equation (\ref{eq:summaryeq}).
It is found that the tensile strength is determined by monomer-connection breaking.
This is consistent with the fact that $P_{x,\mathrm{max}}$ is proportional to $F_\mathrm{c}$ as shown in Equation (\ref{eq:summaryeq}).

\item It is confirmed that the energy dissipation during a stretching simulation support the dependence in Equation (\ref{eq:summaryeq}).
The energy dissipation caused by monomer-connection breaking arises when the tensile stress has a maximum value.
Also, the dependence on the initial volume filling factor corresponds to the fractal dimension of BCCAs, which is about 1.9 \citep{Mukai1992,Okuzumi2009dustagg}.

\item Equation (\ref{eq:summaryeq}) is consistent with the previous experimental \citep{Blum2004,Blum2006,Gundlach2018} and numerical \citep{Seizinger2013} studies of silicate dust aggregates, while it is inconsistent with the previous experimental study of ice dust aggregates \citep{Gundlach2018}.

\item We estimated that the monomer radius of comet 67P has to be 3.3--220 $\mathrm{\mu m}$ using Equation (\ref{eq:summaryeq}).
Assuming that the main material of 67P is silicate and its volume filling factor is about 0.20, we obtained the monomer radius to reproduce its tensile strength of 3--200 Pa.
\end{itemize}

From the point of view of planet formation, the conclusion that the monomer radius of 67P has to be 3.3--220 $\mathrm{\mu m}$ is inconsistent with the typical radius of dust monomers in the interstellar medium: sub-$\mathrm{\mu m}$.
To reduce the monomer radius of 67P, other mechanisms, such as sintering \citep[e.g.,][]{Sirono2017}, to decrease the tensile strength of dust aggregates are needed.

\acknowledgments

We thank Satoshi Okuzumi for fruitful discussions.

\bibliography{paper}

\begin{thebibliography}{}
\expandafter\ifx\csname natexlab\endcsname\relax\def\natexlab#1{#1}\fi
\providecommand{\url}[1]{\href{#1}{#1}}

\bibitem[{{Basilevsky} {et~al.}(2016){Basilevsky}, {Krasil'nikov}, {Shiryaev},
  {Mall}, {Keller}, {Skorov}, {Mottola}, \& {Hviid}}]{Basilevsky2016}
{Basilevsky}, A.~T., {Krasil'nikov}, S.~S., {Shiryaev}, A.~A., {et~al.} 2016,
  Solar System Research, 50, 225

\bibitem[{{Blum} {et~al.}(2014){Blum}, {Gundlach}, {M{\"u}hle}, \&
  {Trigo-Rodriguez}}]{Blum2014}
{Blum}, J., {Gundlach}, B., {M{\"u}hle}, S., \& {Trigo-Rodriguez}, J.~M. 2014,
  \icarus, 235, 156

\bibitem[{{Blum} \& {Schr{\"a}pler}(2004)}]{Blum2004}
{Blum}, J., \& {Schr{\"a}pler}, R. 2004, Physical Review Letters, 93, 115503

\bibitem[{{Blum} {et~al.}(2006){Blum}, {Schr{\"a}pler}, {Davidsson}, \&
  {Trigo-Rodr{\'{\i}}guez}}]{Blum2006}
{Blum}, J., {Schr{\"a}pler}, R., {Davidsson}, B.~J.~R., \&
  {Trigo-Rodr{\'{\i}}guez}, J.~M. 2006, \apj, 652, 1768

\bibitem[{{Blum} \& {Wurm}(2000)}]{Blum2000}
{Blum}, J., \& {Wurm}, G. 2000, \icarus, 143, 138

\bibitem[{{Blum} {et~al.}(2017){Blum}, {Gundlach}, {Krause}, {Fulle},
  {Johansen}, {Agarwal}, {von Borstel}, {Shi}, {Hu}, {Bentley}, {Capaccioni},
  {Colangeli}, {Della Corte}, {Fougere}, {Green}, {Ivanovski}, {Mannel},
  {Merouane}, {Migliorini}, {Rotundi}, {Schmied}, \& {Snodgrass}}]{Blum2017}
{Blum}, J., {Gundlach}, B., {Krause}, M., {et~al.} 2017, \mnras, 469, S755

\bibitem[{{Dominik} \& {Tielens}(1995)}]{Dominik1995}
{Dominik}, C., \& {Tielens}, A.~G.~G.~M. 1995, Philosophical Magazine, Part A,
  72, 783

\bibitem[{{Dominik} \& {Tielens}(1996)}]{Dominik1996}
---. 1996, Philosophical Magazine, Part A, 73, 1279

\bibitem[{{Dominik} \& {Tielens}(1997)}]{Dominik1997}
---. 1997, \apj, 480, 647

\bibitem[{{Fornasier} {et~al.}(2015){Fornasier}, {Hasselmann}, {Barucci},
  {Feller}, {Besse}, {Leyrat}, {Lara}, {Gutierrez}, {Oklay}, {Tubiana},
  {Scholten}, {Sierks}, {Barbieri}, {Lamy}, {Rodrigo}, {Koschny}, {Rickman},
  {Keller}, {Agarwal}, {A'Hearn}, {Bertaux}, {Bertini}, {Cremonese}, {Da
  Deppo}, {Davidsson}, {Debei}, {De Cecco}, {Fulle}, {Groussin}, {G{\"u}ttler},
  {Hviid}, {Ip}, {Jorda}, {Knollenberg}, {Kovacs}, {Kramm}, {K{\"u}hrt},
  {K{\"u}ppers}, {La Forgia}, {Lazzarin}, {Lopez Moreno}, {Marzari}, {Matz},
  {Michalik}, {Moreno}, {Mottola}, {Naletto}, {Pajola}, {Pommerol}, {Preusker},
  {Shi}, {Snodgrass}, {Thomas}, \& {Vincent}}]{Fornasier2015}
{Fornasier}, S., {Hasselmann}, P.~H., {Barucci}, M.~A., {et~al.} 2015, \aap,
  583, A30

\bibitem[{{Fulle} {et~al.}(2016){Fulle}, {Altobelli}, {Buratti}, {Choukroun},
  {Fulchignoni}, {Gr{\"u}n}, {Taylor}, \& {Weissman}}]{Fulle2016}
{Fulle}, M., {Altobelli}, N., {Buratti}, B., {et~al.} 2016, \mnras, 462, S2

\bibitem[{{Goldreich} \& {Ward}(1973)}]{Goldreich1973}
{Goldreich}, P., \& {Ward}, W.~R. 1973, \apj, 183, 1051

\bibitem[{{Greenwood} \& {Johnson}(2006)}]{Greenwood2006}
{Greenwood}, J.~A., \& {Johnson}, K.~L. 2006, Journal of Colloid and Interface
  Science, 296, 284

\bibitem[{{Groussin} {et~al.}(2015){Groussin}, {Jorda}, {Auger}, {K{\"u}hrt},
  {Gaskell}, {Capanna}, {Scholten}, {Preusker}, {Lamy}, {Hviid}, {Knollenberg},
  {Keller}, {Huettig}, {Sierks}, {Barbieri}, {Rodrigo}, {Koschny}, {Rickman},
  {A'Hearn}, {Agarwal}, {Barucci}, {Bertaux}, {Bertini}, {Boudreault},
  {Cremonese}, {Da Deppo}, {Davidsson}, {Debei}, {De Cecco}, {El-Maarry},
  {Fornasier}, {Fulle}, {Guti{\'e}rrez}, {G{\"u}ttler}, {Ip}, {Kramm},
  {K{\"u}ppers}, {Lazzarin}, {Lara}, {Lopez Moreno}, {Marchi}, {Marzari},
  {Massironi}, {Michalik}, {Naletto}, {Oklay}, {Pommerol}, {Pajola}, {Thomas},
  {Toth}, {Tubiana}, \& {Vincent}}]{Groussin2015}
{Groussin}, O., {Jorda}, L., {Auger}, A.-T., {et~al.} 2015, \aap, 583, A32

\bibitem[{{Gundlach} \& {Blum}(2015)}]{Gundlach2015}
{Gundlach}, B., \& {Blum}, J. 2015, \apj, 798, 34

\bibitem[{{Gundlach} {et~al.}(2018){Gundlach}, {Schmidt}, {Kreuzig},
  {Bischoff}, {Rezaei}, {Kothe}, {Blum}, {Grzesik}, \& {Stoll}}]{Gundlach2018}
{Gundlach}, B., {Schmidt}, K.~P., {Kreuzig}, C., {et~al.} 2018, \mnras, 479,
  1273

\bibitem[{Heim {et~al.}(1999)Heim, Blum, Preuss, \& Butt}]{Heim1999}
Heim, L.-O., Blum, J., Preuss, M., \& Butt, H.-J. 1999, Phys. Rev. Lett., 83,
  3328

\bibitem[{{Hirabayashi} {et~al.}(2016){Hirabayashi}, {Scheeres}, {Chesley},
  {Marchi}, {McMahon}, {Steckloff}, {Mottola}, {Naidu}, \&
  {Bowling}}]{Hirabayashi2016}
{Hirabayashi}, M., {Scheeres}, D.~J., {Chesley}, S.~R., {et~al.} 2016, \nat,
  534, 352

\bibitem[{{Israelachvili}(1992)}]{Israelachvili1992}
{Israelachvili}, J.~N. 1992, Surface Science Reports, 14, 109

\bibitem[{{Johansen} {et~al.}(2011){Johansen}, {Klahr}, \&
  {Henning}}]{Johansen2011}
{Johansen}, A., {Klahr}, H., \& {Henning}, T. 2011, \aap, 529, A62

\bibitem[{{Johansen} {et~al.}(2007){Johansen}, {Oishi}, {Mac Low}, {Klahr},
  {Henning}, \& {Youdin}}]{Johansen2007}
{Johansen}, A., {Oishi}, J.~S., {Mac Low}, M.-M., {et~al.} 2007, \nat, 448,
  1022

\bibitem[{{Johnson} {et~al.}(1971){Johnson}, {Kendall}, \&
  {Roberts}}]{Johnson1971}
{Johnson}, K.~L., {Kendall}, K., \& {Roberts}, A.~D. 1971, Proceedings of the
  Royal Society of London Series A, 324, 301

\bibitem[{{Kataoka} {et~al.}(2013{\natexlab{a}}){Kataoka}, {Tanaka}, {Okuzumi},
  \& {Wada}}]{Kataoka2013L}
{Kataoka}, A., {Tanaka}, H., {Okuzumi}, S., \& {Wada}, K. 2013{\natexlab{a}},
  \aap, 557, L4

\bibitem[{{Kataoka} {et~al.}(2013{\natexlab{b}}){Kataoka}, {Tanaka}, {Okuzumi},
  \& {Wada}}]{Kataoka2013}
---. 2013{\natexlab{b}}, \aap, 554, A4

\bibitem[{{Kofman} {et~al.}(2015){Kofman}, {Herique}, {Barbin}, {Barriot},
  {Ciarletti}, {Clifford}, {Edenhofer}, {Elachi}, {Eyraud}, {Goutail}, {Heggy},
  {Jorda}, {Lasue}, {Levasseur-Regourd}, {Nielsen}, {Pasquero}, {Preusker},
  {Puget}, {Plettemeier}, {Rogez}, {Sierks}, {Statz}, {Svedhem}, {Williams},
  {Zine}, \& {Van Zyl}}]{Kofman2015}
{Kofman}, W., {Herique}, A., {Barbin}, Y., {et~al.} 2015, Science, 349,
  doi:10.1126/science.aab0639

\bibitem[{{Kozasa} {et~al.}(1992){Kozasa}, {Blum}, \& {Mukai}}]{Kozasa1992}
{Kozasa}, T., {Blum}, J., \& {Mukai}, T. 1992, \aap, 263, 423

\bibitem[{{Li} \& {Wong}(2013)}]{Li2013}
{Li}, D., \& {Wong}, L.~N.~Y. 2013, Rock Mechanics and Rock Engineering, 46,
  269

\bibitem[{{Meisner} {et~al.}(2012){Meisner}, {Wurm}, \& {Teiser}}]{Meisner2012}
{Meisner}, T., {Wurm}, G., \& {Teiser}, J. 2012, \aap, 544, A138

\bibitem[{{Mukai} {et~al.}(1992){Mukai}, {Ishimoto}, {Kozasa}, {Blum}, \&
  {Greenberg}}]{Mukai1992}
{Mukai}, T., {Ishimoto}, H., {Kozasa}, T., {Blum}, J., \& {Greenberg}, J.~M.
  1992, \aap, 262, 315

\bibitem[{{Okuzumi} {et~al.}(2012){Okuzumi}, {Tanaka}, {Kobayashi}, \&
  {Wada}}]{Okuzumi2012}
{Okuzumi}, S., {Tanaka}, H., {Kobayashi}, H., \& {Wada}, K. 2012, \apj, 752,
  106

\bibitem[{{Okuzumi} {et~al.}(2009){Okuzumi}, {Tanaka}, \&
  {Sakagami}}]{Okuzumi2009dustagg}
{Okuzumi}, S., {Tanaka}, H., \& {Sakagami}, M.-a. 2009, \apj, 707, 1247

\bibitem[{{Ossenkopf}(1993)}]{Ossenkopf1993}
{Ossenkopf}, V. 1993, \aap, 280, 617

\bibitem[{{Paszun} \& {Dominik}(2008)}]{Paszun2008}
{Paszun}, D., \& {Dominik}, C. 2008, \aap, 484, 859

\bibitem[{{P{\"a}tzold} {et~al.}(2016){P{\"a}tzold}, {Andert}, {Hahn}, {Asmar},
  {Barriot}, {Bird}, {H{\"a}usler}, {Peter}, {Tellmann}, {Gr{\"u}n},
  {Weissman}, {Sierks}, {Jorda}, {Gaskell}, {Preusker}, \&
  {Scholten}}]{Patzold2016}
{P{\"a}tzold}, M., {Andert}, T., {Hahn}, M., {et~al.} 2016, \nat, 530, 63

\bibitem[{{Seizinger} {et~al.}(2012){Seizinger}, {Speith}, \&
  {Kley}}]{Seizinger2012}
{Seizinger}, A., {Speith}, R., \& {Kley}, W. 2012, \aap, 541, A59

\bibitem[{{Seizinger} {et~al.}(2013){Seizinger}, {Speith}, \&
  {Kley}}]{Seizinger2013}
---. 2013, \aap, 559, A19

\bibitem[{{Sirono} \& {Ueno}(2017)}]{Sirono2017}
{Sirono}, S.-i., \& {Ueno}, H. 2017, \apj, 841, 36

\bibitem[{{Suyama} {et~al.}(2008){Suyama}, {Wada}, \& {Tanaka}}]{Suyama2008}
{Suyama}, T., {Wada}, K., \& {Tanaka}, H. 2008, \apj, 684, 1310

\bibitem[{{Tanaka} {et~al.}(2012){Tanaka}, {Wada}, {Suyama}, \&
  {Okuzumi}}]{Tanaka2012}
{Tanaka}, H., {Wada}, K., {Suyama}, T., \& {Okuzumi}, S. 2012, Progress of
  Theoretical Physics Supplement, 195, 101

\bibitem[{{Wada} {et~al.}(2007){Wada}, {Tanaka}, {Suyama}, {Kimura}, \&
  {Yamamoto}}]{Wada2007}
{Wada}, K., {Tanaka}, H., {Suyama}, T., {Kimura}, H., \& {Yamamoto}, T. 2007,
  \apj, 661, 320

\bibitem[{{Wada} {et~al.}(2008){Wada}, {Tanaka}, {Suyama}, {Kimura}, \&
  {Yamamoto}}]{Wada2008}
---. 2008, \apj, 677, 1296

\bibitem[{{Youdin} \& {Goodman}(2005)}]{Youdin2005}
{Youdin}, A.~N., \& {Goodman}, J. 2005, \apj, 620, 459

\end{thebibliography}

\end{document}